\documentclass{emulateapj}
\usepackage{apjfonts}

\newcommand\alpharad{\alpha_{\mathrm{rad}}}

\newcommand\Jy{\;\mbox{Jy}}

\newcommand\WHz{\;\mbox{W}\,\mbox{Hz}^{-1}}

\shortauthors{M.~Cara~\& M.~L.~Lister}
\shorttitle{Avoiding spurious breaks in binned LFs}
\begin{document}

\slugcomment{Accepted for publication in the Astrophysical Journal}
\title{Avoiding spurious breaks in binned luminosity functions}

\author{Mihai Cara}
\affil{Department of Physics, Purdue University, 525 Northwestern Avenue, West Lafayette, IN 47907\\E-mail: mcara@physics.purdue.edu}

\and

\author{Matthew L. Lister}
\affil{Department of Physics, Purdue University, 525 Northwestern Avenue, West Lafayette, IN 47907\\E-mail: mlister@physics.purdue.edu}

\begin{abstract}

We show that using either the method of Page~\& Carrera or the well-known $1/V_{\mathrm{a}}$ method to construct the binned luminosity function (LF) of a flux limited sample of Active Galactic Nuclei (AGN) can produce an artificial flattening (or steepening in the case of negative evolution) of the binned LF for bins intersected by the flux cutoff of the sample. This effect is more pronounced for samples with steep and strongly evolving parent LFs but is still present even for non-evolving LFs. As a result of this distortion of the true LF, fitting a model LF to binned data may lead to errors in the estimation of the parameters and may even prompt the erroneous use of broken power law functions. We compute the expected positions of apparent breaks in the binned LF. We show that these spurious breaks in the binned LFs can be avoided if the binning is done in the flux--redshift plane instead of the typically used luminosity--redshift plane. Binning in the flux--redshift plane can be used in conjunction with the binning in the luminosity--redshift plane to test for real breaks in the binned LFs and to identify the features that are the result of binning biases. We illustrate this effect for most typical forms of luminosity dependence and redshift evolution and show how the proposed method helps address this problem. We also apply this method to the MOJAVE AGN sample and show that it eliminates an apparent break in the binned LF.

\end{abstract}


\section{Introduction}

The luminosity function (LF) of active galactic nuclei (AGN) and its redshift dependence is highly useful for studying the cosmological evolution of AGNs
and for statistical tests of AGN unification models. Typically the binned LF is obtained using the standard $1/V_{\max}$ method \citep[][]{Schmidt68,Felten76} or its generalization for combined samples $1/V_{\mathrm{a}}$ \citep[][]{Avni80}. The evolution of the LF is estimated using the $\left\langle V/V_{\max} \right\rangle$ method \citep[][]{Schmidt68}. Sometimes a model LF is derived by fitting a model LF to the binned LF \citep[see, e.g.,][]{Padovani07}. However, the binned LF obtained using the $1/V_{\mathrm{a}}$ method suffers from several biases \citep[see][and references therein]{LaFranca97,Wisotzki98} and for this reason many prefer fitting the model LF to the unbinned data with, e.g., maximum likelihood techniques. In this case the  binned LF, however, is often used before performing the fit to observe the overall behavior of the luminosity function, e.g., to decide its evolution type and to decide whether or not a double power law LF should be fitted to the data instead of a single power law. It can also be used after the fit to visualize the goodness of fit.

\citet{PC00} described a different method of computing the binned LF that, according to the authors, gives better estimates of the LF for lower luminosity bins that are close to the flux limit of the survey using an improved (over $1/V_{\mathrm{a}}$) estimator. However, their method does not eliminate all the biases associated with undersampled bins that are near the flux cutoff of the survey \citep[see, e.g.,][]{Miyaji01}. In a recent paper \citep{Cara08} we derived the radio luminosity function of the MOJAVE survey of core-selected AGN \citep{LH05}. While a single power law provided a good fit to the unbinned data, the binned LF, computed using the method of \citet{PC00}, was flatter at lower luminosities, and suggested a double power law LF \citep[see, e.g., ][]{Arsh06}. After an analysis of the problem \citep[see][]{Cara08} we concluded that the flattening of the binned LF was due to the fact that the values of the LF computed over small chunks of undersampled (because of the flux cutoff) bins are not good approximations of the value of the LF at the centers of the bins, particularly when the LF is a steep and strongly evolving function across the bin.

There are many other examples in the literature where binned luminosity functions have been extended well below the flux cutoff of the survey. For example, \citet[][Fig. 10]{Dunlop90} used the binned LF to compare the prediction of their pure luminosity evolution (PLE) models with the data. The plot displays a clear flattening of the LF for bins below the flux cutoff. Because of the multiple flux limits in their data, this flattening happens gradually and, therefore, the binned LFs in that plot are better described as being ``bent'' than ``broken''.

A somewhat different example of using bins below the flux cutoff comes from \citet{Urry95}. In Figure 14 the authors present the \emph{local} LFs of the FR~II galaxies, steep spectrum radio quasars (SSRQ) and flat spectrum radio quasars (FSRQ). In this case, because the luminosities have been de-evolved, the LFs have been constructed essentially using one large redshift bin with a modified (de-evolved) flux limit. Therefore, almost all the luminosity bins presented in Figure 14 of \citet{Urry95} lie below the survey limit (however, the precise number of bins below the limit is difficult to determine because the redshift bin width is not presented). At very low luminosities, particularly at luminosities where the survey limit starts cutting into the number of objects in a bin, the number densities will start decreasing because the incomplete bins, due to the shape of the flux cutoff, will preferentially sample a different part of the LF than would be the case for a complete bin. We will show in Section~\ref{sec:Method} that the position of the dominant break will depend on the minimum redshift of the bin in the case of a non-evolving luminosity function.

A final example is that of the binned luminosity function of the quarter-Jansky sample from \citet{Wall05}. The flattening of the LF of this sample at low luminosities produces the illusion of the luminosity evolution, which was recognized by the authors as such. The authors attributed the flattening of the LF at lower luminosities to an intrinsic spread in radio spectral indices.

Since this flattening (or steepening in the case of negative evolution) is caused by undersampling of the bins cut by the survey limit, one solution to eliminate these spurious breaks is to stop binning as soon as the lowest luminosity bin reaches the survey limit. This, however, can lead to the exclusion of data points lying between this last complete bin and the survey limit. \citet{LaFranca97,Miyaji01} proposed to correct for biases in the undersampled bins by prorating them for the missing parts using a model LF determined by independent means. This approach, however, has the disadvantage of depending on an assumed model. \citet{Efstathiou88} have developed a non-parametric ``step-wise maximum likelihood'' method that allows a binned LF to be fit directly to the unbinned data. While these methods their own advantages and disadvantages, we have chosen to focus in this paper on the method of Page~\& Carrera. We note that any method that computes the values of the LF over incomplete bins must, in one way or another, use some sort of extrapolation to the parts of the bins lying below the flux limit.

In this paper we show that the biases associated with undersampled bins can produce artificial breaks in the derived luminosity functions, particularly in the case of strong evolution. As evolution of the LF is typically judged by the shifts of features in the LF, these artificial breaks, when interpreted as real, can produce the illusion of luminosity evolution. We show that the flattening (or breaking) of the binned LF for bins close to the flux cutoff of the survey can be minimized if the binning is performed in the flux--redshift plane instead of the luminosity--redshift plane as it is usually done. The improvement is most apparent at lower redshifts where, due to the steepness of the luminosity cutoff in the $\log L-z$ space, many bins are undersampled, and therefore give strongly biased estimates of the luminosity function. While our analysis of the distortion of the shape of the binned LFs is performed using the method of \citet{PC00} for constructing binned luminosity functions, this analysis equally applies to the $1/V_{\mathrm{a}}$ method as well. The binned LFs constructed using the $1/V_{\mathrm{a}}$ method will exhibit larger deviations from the true LF than the method of \citet{PC00} at extremely low luminosities ($L<L_{\mathrm{break,3}}$, see equation~\ref{eqn:PosBreak3}). However, in the case of non-evolving LFs, the $1/V_{\mathrm{a}}$ method, unlike the method of Page~\& Carrera, will not distort the shape of the LF for $L_{\mathrm{break,3}}<L<L_{\mathrm{break,1}}$ because it assigns a variable weight to each object in the bin and can correctly handle the flux limit down to lower luminosities~($L_{\mathrm{break,3}}$). Regardless of the method used, any determination of the LF below $L_{\mathrm{break,3}}$ is pure extrapolation and should be avoided.

The outline of the paper is as follows: In \S~\ref{sec:Method} we describe both binning methods and discuss the main advantages of the binning in the flux--redshift plane over the traditional luminosity--redshift plane binning method. In \S~\ref{sec:MCsim} we apply these two binning methods to several model luminosity functions and show that the LF binned in the flux--redshift plane is a better approximation to the model LF. We construct the binned LF of the MOJAVE sample using both methods in \S~\ref{sec:MOJAVELF}. We summarize our findings and present our conclusions in \S~\ref{sec:conclusions}.

Throughout this paper we assume a cosmology with $\Omega_m=0.3$, $\Omega_{\Lambda}=0.7$, $\Omega_{r}=0$ and $H_0=70\,\mbox{km}\,\mbox{s}^{-1}\,\mbox{Mpc}^{-1}$. All luminosities are quoted as monochromatic luminosities at specific frequency $\nu$.  We also adopt the following convention for the spectral index, $\alpharad$: $S_{\nu}\propto\nu^{\alpharad}$ and assume that $\alpharad=0$ for all model LFs and plots in this paper.

\section{Method}\label{sec:Method}

The differential luminosity function of a sample of astrophysical objects is defined as the number of objects per unit
comoving volume per unit luminosity interval:
\begin{equation}\label{eqn:defLF}
\phi \left(L,z\right)=\frac{d^2 N}{dV dL}
\end{equation}
but, because of the typically large span of the luminosities, it is often defined in terms of $\log L$:
\begin{equation}\label{eqn:defLFlog}
\phi \left(\log L,z\right)=\frac{d^2 N}{dV d\log L}
\end{equation}
and the relationship between the two forms is given by:
\begin{equation}\label{eqn:LFlogLF}
\phi \left(\log L,z\right)=\ln 10\; L\; \phi \left(L,z\right).
\end{equation}
One of the advantages of working with the LF defined in terms of $\log L$ is that for the most common forms of the LF used in model fitting, i.e., the simple power law
\begin{equation}\label{eqn:singlepow}
\phi \left(L,z\right)=\frac{\phi_0}{L_{*}}\left( \frac{L}{L_{*}}\right)^{\alpha},
\end{equation}
the plot of $\log \phi$ vs. $\log L$ will show a simple linear dependence:
\begin{equation}\label{eqn:singlepowloglog}
\log \phi \left(\log L,z\right)=\left( \alpha + 1 \right) \log L + \log \left( \ln 10 \right)-\left( \alpha + 1 \right) \log L_*
\end{equation}
with a slope $\left( \alpha + 1 \right)$. In equations (\ref{eqn:singlepow}) and (\ref{eqn:singlepowloglog}) $L_*$ is an arbitrary positive constant.

\begin{figure}[htb]
\begin{center}
\includegraphics[angle=-90, scale=0.315]{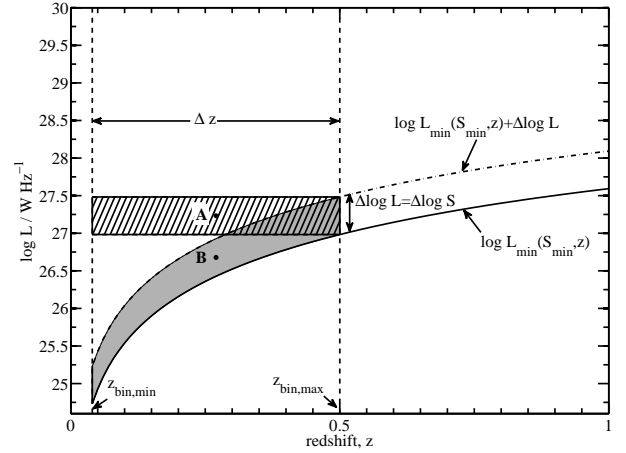}
\end{center}
\caption{\small A typical bin (``A'') in $\log L-z$ space is shown together with a bin (``B'') that is a rectangle in $\log S-z$ space. Both rectangles lie just above the flux limit of the survey (solid line $L_{\min}\left( S_{\min},z \right)$). The bins' height is $\Delta \log L=\Delta \log S = 0.5$ and they span a redshift range $0.04<z<0.5$. The dash-dotted line shows the upper limit of the bin in the $\log S-z$ plane (line  $L_{\min}\left( S_{\min},z \right)+\Delta \log L$). The bin ``A'' is centered on $\log L_{\mathrm{break,1}}\approx 27.23$ and spans a range of luminosities $26.98<\log L<27.48$. The bin corresponding to $\log L_{\mathrm{break,2}}\approx 26.73$ that would span a range of luminosities $26.48<\log L<26.98$ (i.e., it would be adjacent to bin ``A'') and the bin corresponding to $\log L_{\mathrm{break,3}}\approx 24.98$ that would span a range of luminosities $24.73<\log L<25.23$ are omitted here for clarity. }
\label{fig:LzLsBinning}
\end{figure}

The usual way to construct the binned LF of a sample of objects is to divide the $\log L-z$ plane into luminosity--redshift bins and to estimate the space density in a bin using, for example, the $1/V_{\max}$ method or other methods. In this paper we use the improved method of \citet{PC00}, that is, we assume that the LF at the center of a bin with a luminosity interval $\log L_{\min}$ and $\log L_{\max}$ and a redshift interval $z_{\min}$ and $z_{\max}$ can be approximated as:
\begin{equation}\label{eqn:BinLFestLz}
\phi \left(\log L_\mathrm{c},z_\mathrm{c} \right) \approx
   \frac{N}{\int_{z_{\min}}^{z_{\max}}
   \int_{\log L_{\min}(z)}^{\log L_{\max}}\frac{dV}{dz}\,dz\,d\log L},
\end{equation}
where $N$ is the number of sources detected within the bin limits and $L_{\min}(z)$ is the minimum luminosity within
the bin at which we can still detect an object. In Figure~\ref{fig:LzLsBinning} we show a luminosity-redshift bin (hatched bin ``A'') positioned so that it just touches the flux cutoff of the survey at $z=0.5$. Any bin of the same luminosity width $\Delta \log L$ sampled in the $\log L-z$ plane and centered below the center of the bin ``A'' will have some parts of it lying below the survey's flux limit. If the luminosity function of the parent sample is steep and strongly evolving across the bin, then equation~(\ref{eqn:BinLFestLz}) will produce stronger biases for undersampled bins than for bins completely above the flux cutoff. We note that the shape of the binned luminosity function will not be affected by biases as long as these biases are constant for all bins in a given redshift interval. However, this is not the case for the undersampled bins: given a negative power law index for the LF in the parent sample ($\alpha < 0$), in the case of positive evolution, the change in bias will tend to flatten the LF, while in the case of negative evolution, it will tend to steepen the LF at luminosities near and below the luminosity limit. The position of this break, using $L=4\pi S D_{\mathrm{L}}^2(z)(1+z)^{-(1+\alpharad)}$, will be given by
\begin{equation}\label{eqn:PosBreak1}
\log L_{\mathrm{break,1}}=\log\left[\frac{4\pi S_{\min} D_{\mathrm{L}}^2(z_{\max})}{(1+z_{\max})^{1+\alpharad}}\right]+\Delta\log L / 2,
\end{equation}
where $S_{\min}$ is the flux limit of the survey and $D_{\mathrm{L}}(z)$ is the luminosity distance. This break luminosity corresponds to the luminosity coordinate of the center of the bin ``A'' shown in Figure~\ref{fig:LzLsBinning}. The break (i.e., the difference between the slopes before and after the break luminosity) will be more pronounced for more strongly evolving LFs. Another change in shape of the LF is possible when the flux cutoff line passes through the upper-right (high redshift and high luminosity) corner of the bin:
\begin{equation}\label{eqn:PosBreak2}
\log L_{\mathrm{break,2}}=\log L_{\mathrm{break,1}}-\Delta\log L .
\end{equation}
Finally, the last break will occur around 
\begin{equation}\label{eqn:PosBreak3}
\log L_{\mathrm{break,3}} = \log\left[\frac{4\pi S_{\min} D_{\mathrm{L}}^2(z_{\min})}{(1+z_{\min})^{1+\alpharad}}\right]+\Delta\log L / 2,
\end{equation}
which is the coordinate of the center of the bin whose lower-left (low redshift and low luminosity) corner is just touching the survey limit. For non-evolving LFs, this is the dominant break, since at luminosities lower than  $L_{\mathrm{break,3}}$ we will be sampling only the lower density part of the bin. The coordinates of the last two breaks are approximate because their effective position will depend on the bin dimensions, steepness of the survey limit within the bin limits and steepness of the LF itself. In addition, if $\alpharad$ is not constant, the positions of the above breaks will be blurred since the breaks will occur at different luminosities for sources with different spectral indexes. Alternatively, one can think that each source has its own survey limit \citep[similar to the ``single source survey'' in][]{Wall05} and the observed break will be the result of overlapping of all these individual breaks.

In the case of samples with multiple limits, the breaks given by equations (\ref{eqn:PosBreak1}), (\ref{eqn:PosBreak2}) and (\ref{eqn:PosBreak3}) should be computed for each flux limit. That is, the total number of breaks in the binned (in $\log L-z$ plane) LF potentially will be equal to three times the number of flux limits. The strengths of each break will depend on the sky area covered for each flux limit (i.e., by the contribution of each subsample). The first break, however, will occur at $L=L_{\mathrm{break,1}}$ computed for the largest flux limit in the sample.

The undersampled bins distort the shape of the LF and in principle should not be used in the analysis. Doing this, however, would discard a significant amount of information since, in samples with $\alpha < 0$, sources tend to accumulate closer to the luminosity limit of the survey. The loss of information, however, can be avoided if the sampling is done in the flux--redshift plane while keeping all bins above the flux cutoff and thus avoiding the distortion of the LF shape. Sampling in the $\log S-z$ plane was used previously by \citet{Kochanek96} to minimize the effects of cosmology and variations in the spectral index. Here we will convert the binned LF back to $\log L-z$ coordinates to compare the shapes of the LFs obtained through both methods. The method of constructing the binned LF in the $\log S-z$ space is essentially similar to the method described above. Noting that for constant spectral indexes $d\log S = d\log L$ and that the limits of integration in the equation~(\ref{eqn:BinLFestLz}) are fixed in the $\log S-z$ space, the double integral can be easily computed and the value of the binned LF can be estimated as
\begin{equation}\label{eqn:BinLFestSz}
\phi \left(\log S_\mathrm{c},z_\mathrm{c} \right) \approx
   \frac{N}{\left( V\left( z_{\mathrm{\max}}\right)-V\left( z_{\mathrm{\min}}\right)\right)\Delta\log S},
\end{equation}
where $z_{\mathrm{\min}}$ and $z_{\mathrm{\max}}$ are the redshift limits of the bin, and $V(z)$ is comoving volume.

We have mentioned above that the distortion of the LF is strongest at undersampled bins, but even the complete bins are not fully immune from distortions. However, for complete bins, we expect the bias, i.e., the difference between the true value of the LF at the center of the bin and the value computed using either equation~(\ref{eqn:BinLFestLz}) or (\ref{eqn:BinLFestSz}), to change little, at least for well behaved parent LFs such as the single power law LF from equation~(\ref{eqn:singlepow}) or a smoothed two power law LF of the form
\begin{equation}\label{eqn:dblpow}
\phi \left(L,z\right)=\frac{\phi_0/L_{\mathrm{break}}}{\left( \frac{L}{L_{\mathrm{break}}}\right)^{-\alpha_1}+\left( \frac{L}{L_{\mathrm{break}}}\right)^{-\alpha_2}},
\end{equation}
at least for pure density evolution (PDE). In the case of pure luminosity evolution (PLE) the two power law LF may exhibit more complex biases and possible larger distortions, but these should be confined mainly to the region around the break luminosity $L_{\mathrm{break}}$.

From Figure~\ref{fig:LzLsBinning} we see that the complete bins in the two methods, when centered around the same luminosity, sample different regions of the $\log L-z$ plane and even if the binned LF may be distorted, we would expect these distortions to be different. This can be used to test whether a break seen in a binned LF is real: \emph{a real break should appear with high probability in the binned LFs regardless of the sampling method}, while a spurious break that is due to binning biases will depend on the sampling method.

Another immediate observation from Figure~\ref{fig:LzLsBinning} is that the centers of the complete bins closest to the luminosity limit are lower for bins sampled in the $\log S-z$ plane. It can be shown that for the lowest bins in the two sampling methods (bins ``A'' and ``B'' in Figure~\ref{fig:LzLsBinning})
\begin{eqnarray}
\log L_{\mathrm{c}}^{\mathrm{B}} & = & \log L_{\mathrm{c}}^{\mathrm{A}}-\left(\alpharad+1 \right)\log\left(\frac{1+z_{\max}-\Delta z/2}{1+z_{\max}}\right) \nonumber \\
 & & + 2\log\left(\frac{D_{\mathrm{L}}\left(z_{\max}-\Delta z/2\right)}{D_{\mathrm{L}}\left(z_{\max}\right)}\right)\label{eqn:LcLcDif}
\end{eqnarray}
so that $\log L_{\mathrm{c}}^{\mathrm{B}}$ is smaller than $\log L_{\mathrm{c}}^{\mathrm{A}}$ as long as $D^2_{\mathrm{L}}(z)(1+z)^{-(1+\alpharad)}$ is an increasing function of redshift. Thus, \emph{if we limit ourselves to complete bins only}, the sampling in the $\log S-z$ plane allows the LF to be determined down to lower luminosities than when sampling in the $\log L-z$ plane.

Finally, for steep LFs, the density of the sources will be greater along the flux limit of the survey so that the lower luminosity bins sampled in the $\log S-z$ plane will contain more sources than the bins sampled in the $\log L-z$ plane. This allows one to reduce the luminosity width of the bins and obtain a finer sampling of the LF near the flux limit.

One disadvantage of the sampling in the $\log S-z$ plane is that the bins, when converted to the $\log L-z$ plane, span a larger range of luminosities when compared to direct sampling in the $\log L-z$ plane. This can result in blurring of real features (if present) in the binned LF. This blurring can be somewhat minimized by reducing the redshift width of the bins at the expense of higher Poisson errors in the high luminosity bins.

\section{Comparison of binned and model luminosity functions}\label{sec:MCsim}

In this section we construct the binned LF for several commonly used forms of model luminosity functions and their evolution. We compute the binned LF in the $\log L-z$ plane from equation~(\ref{eqn:BinLFestLz}) with $N$ computed as: 
\begin{equation}\label{eqn:BinLFestLzN}
N_{\mathrm{model}}=\int_{z_{\min}}^{z_{\max}}\int_{\log L_{\min}(z)}^{\log L_{\max}}
\phi(\log L,z)\frac{dV}{dz}\,dz\,d\log L .
\end{equation}
Similarly, for bins sampled in the $\log S-z$ plane we use 
\begin{equation}\label{eqn:BinLFestSzN}
N_{\mathrm{model}}=\int_{z_{\min}}^{z_{\max}}\int_{\log S_{\min}}^{\log S_{\max}}
\phi(\log L(S,z),z)\frac{dV}{dz}\,dz\,d\log S .
\end{equation}
Alternatively, one could use Monte Carlo simulations to produce the desired samples and construct the binned LFs from these simulated samples. However, these samples would need to be very large to minimize the low number statistics at the extreme luminosity ends of the binned LFs. In this section we will assume a survey flux limit $S_{\min}=1.5\Jy$. The parent population has redshift range $z_1=0.05$ and $z_2=4$, and luminosity range from $L_1=5\times10^{24}\WHz$ to $L_2=10^{30}\WHz$. These values are based on \citet{Cara08}. We will normalize the LFs so that $N=1$.

The biases in the binned LFs will be different for the two sampling methods, but since we want to compare the distortions of the LFs, a constant shift of the LF is not important for our purposes. Therefore, in order to make it easier to compare the model LF with the binned LFs we shift the binned LFs by an amount equal to the bias of the first (i.e., lowest luminosity) complete bin, computed as $\left\langle \phi \right\rangle_{\mathrm{bin}}-\left.\phi \right|_{\mathrm{bin}\;\mathrm{center}}$.

\begin{figure}[tb]
\begin{center}
\includegraphics[angle=-90, scale=0.315]{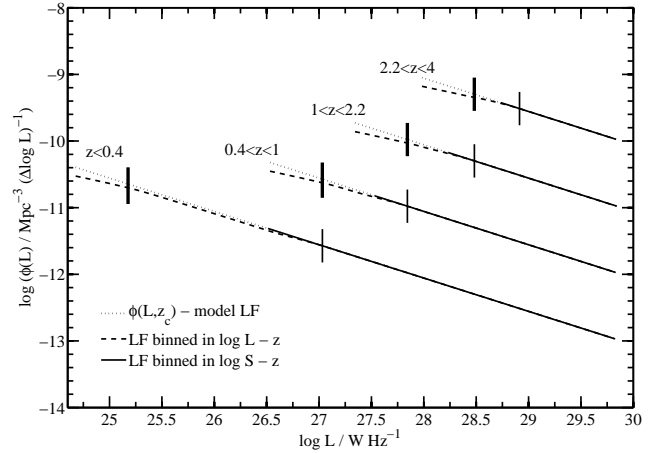}
\end{center}
\caption{\small Differential (dotted line) and binned luminosity functions in $\log L-z$ plane using the method of Page~\& Carrera (dashed line) and in $\log S-z$ plane (continuous line) for a non-evolving single power law LF with $\alpha=-2.5$. The redshift intervals are $0.05<z<0.4$,  $0.4<z<1$, $1<z<2.2$ and $2.2<z<4$. The LFs in higher redshift intervals are shifted upwards along the density axis by 1 dex compared to the LF in the previous redshift interval. The thin vertical lines show the position of the last complete bin ($L_{\mathrm{break,1}}$) in the $\log L-z$ plane and the thick vertical lines show the position of the break $L_{\mathrm{break,3}}$. The solid lines are truncated at luminosities given by equation~(\ref{eqn:LcLcDif}) with $\log L_{\mathrm{c}}^{\mathrm{A}} = L_{\mathrm{break,1}}$, below which the LFs binned in $\log S-z$ plane would be extrapolated. The binned LFs were computed using $\Delta \log L = 0.5$. }
\label{fig:onepowBinnedLFnoev1}
\end{figure}

\begin{figure}[tb]
\begin{center}
\includegraphics[angle=-90, scale=0.315]{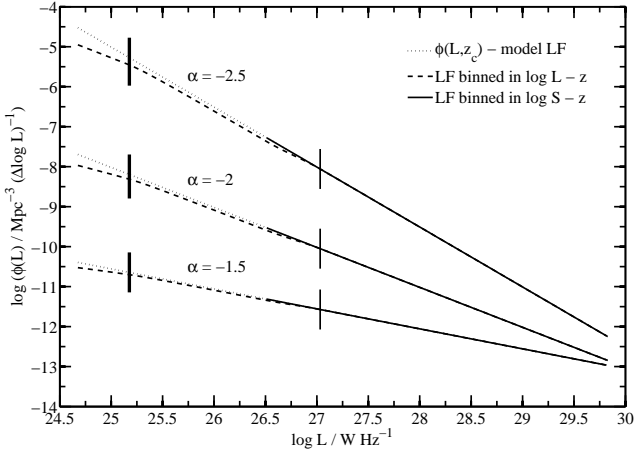}
\end{center}
\caption{\small Differential (dotted line) and binned luminosity functions in $\log L-z$ plane (dashed line) and in $\log S-z$ plane (continuous line) for a single power law LF for three values of the slope $\alpha=-1.5$, $\alpha=-2$ (shifted upwards by 1 dex compared to the $\alpha=-1.5$ LF) and $\alpha=-2.5$ (shifted upwards by 3 dex compared to the $\alpha=-1.5$ LF). The redshift interval is $0.05<z<0.4$. The thin vertical lines show the position of the last complete bin ($L_{\mathrm{break,1}}$) in the $\log L-z$ plane and the thick vertical lines show the position of the break $L_{\mathrm{break,3}}$. The binned LFs were computed using $\Delta \log L = 0.5$. }
\label{fig:onepowBinnedLFnoev2}
\end{figure}

We begin by showing the effects of the binning in $\log L-z$ plane on a moderately steep ($\alpha=-1.5$) single power law LF with no evolution. The binning is performed in four redshift intervals with $\Delta\log L=0.5$. The results are presented in Figure~\ref{fig:onepowBinnedLFnoev1} with the LFs in higher redshift bins shifted upward by 1 dex compared to the LF in the previous bin. The position of the beginning of the break corresponding to $L_{\mathrm{break,1}}$ from equation~(\ref{eqn:PosBreak1}) is marked by a thin vertical line for every LF and the thick vertical lines show the positions of $L_{\mathrm{break,3}}$. The ends of the solid lines correspond to $\log L_{\mathrm{c}}^{\mathrm{B}}$ from equation~(\ref{eqn:LcLcDif}) with $\log L_{\mathrm{c}}^{\mathrm{A}} = L_{\mathrm{break,1}}$ for each redshift bin. Although LFs should never be calculated below $L_{\mathrm{break,3}}$, we plot them here only to illustrate the significant deviations that result. Several observations can be drawn from Figure~\ref{fig:onepowBinnedLFnoev1}: 1) even for a moderately steep non-evolving LF, binning below the survey limit in the $\log L-z$ plane produces a broken LF; 2) the effective position of the break will depend on the redshift limits of the bins, and for a non-evolving LF the break is closer to $L_{\mathrm{break,3}}$ from equation~(\ref{eqn:PosBreak3}); 3) the break shifts along the luminosity axis, thus mimicking luminosity evolution; 4) the LF binned in the $\log S-z$ plane preserves the shape of the model differential LF. While the difference between  the two binned LFs is negligible over the common luminosity range, it becomes more pronounced for steeper and stronger evolving functions. To illustrate the effect of steepening of the LF, in Figure~\ref{fig:onepowBinnedLFnoev2} we plot the binned LFs in the first redshift interval ($0.05<z<0.4$) for three values of the model LF slope: $\alpha=-1.5$,  $\alpha=-2$, and $\alpha=-2.5$. As expected, the break is more evident for steeper LFs. We note here that for non-evolving LFs, the $1/V_\mathrm{a}$ method will correctly reproduce the shape of the LF down to $L_{\mathrm{break,3}}$ because its variable weight can correctly handle the flux cutoff for bins above $L_{\mathrm{break,3}}$, while the method of Page~\& Carrera assumes a constant value of the LF within a bin.

\begin{figure}[tb]
\begin{center}
\includegraphics[angle=-90, scale=0.315]{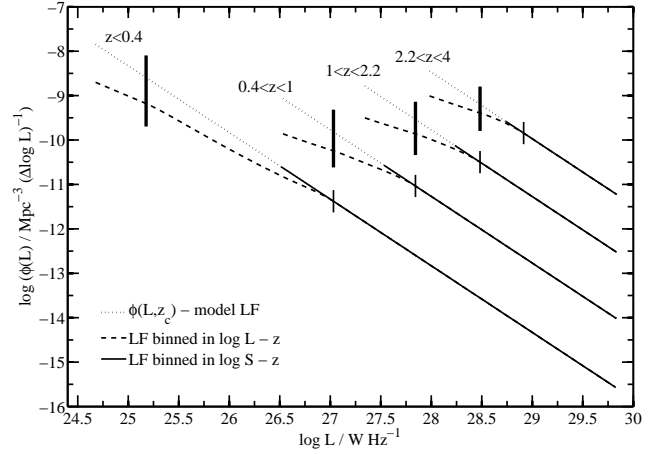}
\end{center}
\caption{\small Same as in Figure~\ref{fig:onepowBinnedLFnoev1}, except the LF has pure density evolution (PDE) of the form $\exp\left[k\,\tau(z)\right]$ with $k=5$. The luminosity slope is $\alpha=-2.5$. }
\label{fig:onepowLFLkbk}
\end{figure}

We now explore the effects of binning on LFs containing pure density evolution. There are several analytic forms of evolution often used in modeling of the luminosity functions: ~$(1+z)^k$, $\exp\left[k \tau(z)\right]$ (where $\tau(z)$ is the look-back time in units of Hubble time) and $\exp\left[-\frac{1}{2}\left(\frac{z-z_0}{\sigma}\right)^2\right]$. The difference between the first two evolution forms is minor for our purposes and we choose the second form of evolution with $k=5$. This value of the evolution parameter is consistent with the value reported by \citet{Padovani07} for the DXRBS flat spectrum radio quasar sample when the latter is converted from PLE to PDE. The binned and differential LFs with exponential density evolution and slope $\alpha=-2.5$ are presented in Figure~\ref{fig:onepowLFLkbk}. Compared to the non-evolving LF, the break in the LF binned in the $\log L-z$ plane is now much more pronounced, while the LF binned in the $\log S-z$ plane does not show any break. Intuitively, this is because the mean redshifts in undersampled bins deviate from the mean redshifts in complete bins by larger values when evolution is present. Again, the shift of this break along the luminosity axis produces the illusion of luminosity evolution.

\begin{figure}[tb]
\begin{center}
\includegraphics[angle=-90, scale=0.315]{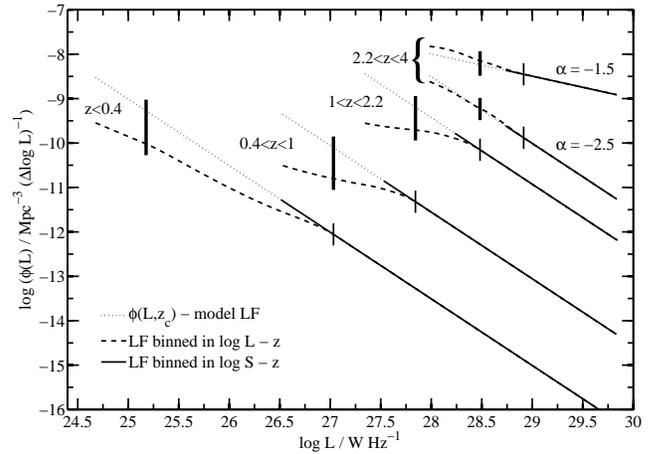}
\end{center}
\caption{\small Same as in Figure~\ref{fig:onepowBinnedLFnoev1} except the LF is evolving with PDE of the form $\exp\left[-\frac{1}{2}\left(\frac{z-z_0}{\sigma}\right)^2\right]$ with $z_0=2.34$ and $\sigma=0.631$. The luminosity slope is $\alpha=-2.5$ for the lower four LFs and $\alpha=-1.5$ for the upper LF. The LFs in higher redshift intervals a shifted upwards along the density axis by 1 dex compared to the LF in the previous redshift interval and the LF in the $2.2<z<4$ with $\alpha=-1.5$ was shifted upwards one more dex compared to the LF with $\alpha=-2.5$ slope in the same redshift interval. The thin vertical lines show the position of the last complete bin ($L_{\mathrm{break,1}}$) in the $\log L-z$ plane and the thick vertical lines show the position of the break $L_{\mathrm{break,3}}$. }
\label{fig:onepowLFGauss}
\end{figure}

The Gaussian form of evolution is different from the first two forms because it produces strong negative evolution for $z>z_0$. For this form of evolution we use the parameter values for model B in cosmology III from \citet{Willott98}: $z_0=2.34$ and $\sigma=0.631$ with luminosity slope $\alpha=-2.5$. The results are presented in Figure~\ref{fig:onepowLFGauss}. Compared to the previous evolution model in Figure~\ref{fig:onepowLFLkbk}, the apparent breaks are much stronger in the lower redshift bins, however the break almost disappears in the highest redshift interval. This is due to negative evolution for $z>2.34$, which actually increases the slope of the LF in the lower luminosity range. For this particular selection of parameters, the steepening of the LF due to the negative evolution is balanced somewhat by the artificial flattening created by the binning method. To better illustrate this point we replotted the LF in the last redshift interval using a luminosity distribution slope $\alpha = -1.5$. In this case the flattening due to steep luminosity dependence cannot balance completely the steepening due to negative evolution.

If the artificial offsets are removed from the LF curves in Figure~\ref{fig:onepowLFGauss}, one can see that unlike the case of the bias due to spectral spread reported by \citet{Wall05}, the apparent luminosity evolution due to spurious breaks in the binned LFs is created by the breaks at \emph{different} space densities. This is in good agreement with decrease in the density at which the breaks occur seen in Figure 3 of \citet{Wall05}. Also, the steepening below the breaks of the LF in the case of negative evolution is in good agreement with the steepening of the binned LF in the highest redshift interval in Figure 3 of \citet{Wall05}. The negative evolution in that redshift interval is revealed by the shift of this function to the left (towards lower luminosities). We point out that  the break in the LF due to undersampled bins ($L_{\mathrm{break,1}}$) will occur at larger luminosities than the ones corresponding to the vertical bars in Figure 3 of \citet{Wall05}, since their sample has multiple flux limits. As discussed in Section~\ref{sec:Method}, the flattening will begin at $L_{\mathrm{break,1}}$, computed for the largest flux limit of the survey.

\begin{figure}[tb]
\begin{center}
\includegraphics[angle=-90, scale=0.315]{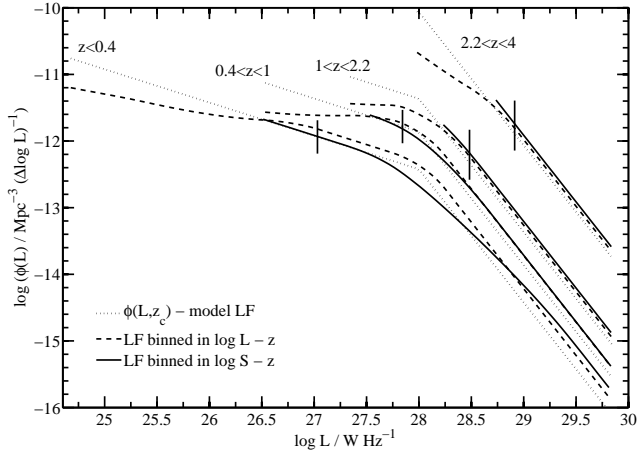}
\end{center}
\caption{\small Luminosity functions for a two power law luminosity distribution with $\alpha_1=-1.5$ and $\alpha_2=-3$ and a sharp break. We assume exponential PDE of the form $\exp\left[k\,\tau(z)\right]$ with $k=5$. Notations are the same as in Figure~\ref{fig:onepowBinnedLFnoev1}. In this plot only the LF in the $2.2<z<4$ redshift interval was shifted upwards by 1 dex. }
\label{fig:twopowBinnedLF}
\end{figure}

Finally, we examine how real features in LFs are blurred by different sampling methods. For this purpose we will use a two power law model LF with a sharp break:
\begin{equation}\label{eqn:sharpdblpow}
\phi \left(L,z\right)=\phi_0
\cases{\left( L/L_{\mathrm{break}}\right)^{\alpha_1}, & $L_1< L < L_{\mathrm{break}}$, \cr
\left( L/L_{\mathrm{break}}\right)^{\alpha_2}, & $L_{\mathrm{break}}< L < L_2$,}
\end{equation}
with $\alpha_1=-1.5$, $\alpha_2=-3$, and $\log L_{\mathrm{break}}=28$ with $\exp\left[5 \tau(z)\right]$ as the parametrized density evolution. The results are presented in Figure~\ref{fig:twopowBinnedLF}. Again, the LFs binned in the $\log L-z$ plane diverge from the differential LF below the luminosity limit and yield a flatter low luminosity slope. Unlike the single power law cases discussed previously, in this case the real break in the LF eliminates the appearance of luminosity evolution. We can see that with binning in the $\log S-z$ plane it is possible to reach lower luminosities than $\log L-z$ binning without producing artificial breaks. However, this comes at the expense of blurring the real breaks: we can see that the (real) break in the LF binned in the $\log S-z$ plane is wider than the (real) break of the LF sampled in the $\log L-z$ plane. This blurring can be minimized somewhat by reducing the redshift width of the bins at the expense of smaller number statistics in the high luminosity bins.

\section{The binned LF of the MOJAVE sample}\label{sec:MOJAVELF}

\begin{figure}[tbp]
\begin{center}
\begin{tabular}{c}
\includegraphics[angle=-90,scale=0.315]{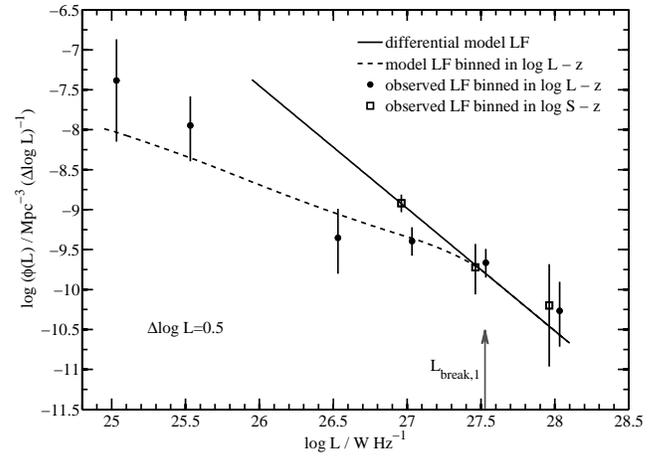} \\
\includegraphics[angle=-90,scale=0.315]{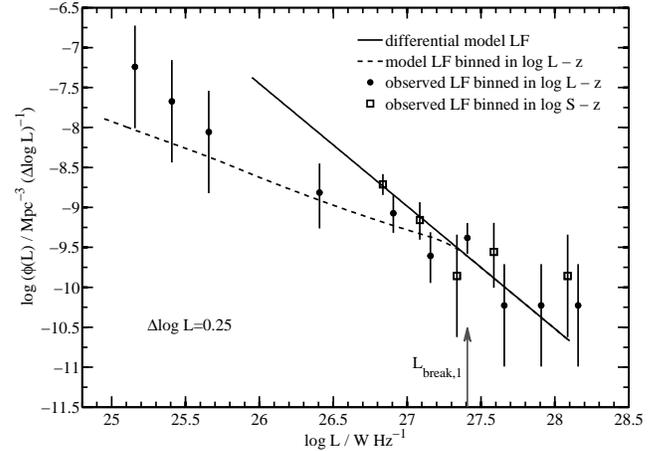}
\end{tabular}
\end{center}
\caption{\small MOJAVE LF of FR~II-only sources for the redshift interval $0.04<z<0.7$ and $\Delta \log L = 0.5$ (upper panel) and $\Delta \log L = 0.25$ (lower panel). The continuous line shows the differential model LF, while the dashed line shows the binned model LF. The filled circles show the observed LF binned in the $\log L-z$ plane and open squares -- the observed LF binned in the $\log S-z$ plane. The gray arrows indicate the position of the lowest luminosity complete bin in the $\log L - z$ plane corresponding to $L_{\mathrm{break,1}}$. }
\label{fig:MOJAVErebin}
\end{figure}

As an illustration using actual data, we construct the binned LF of the MOJAVE sample for the FR~II-only sample \citep{LH05} using binning in the $\log S-z$ plane and binning in $\log L-z$ plane using the method of \citet{PC00}. The LFs constructed in the $\log L-z$ plane were presented in Figure 4 of \citet{Cara08}. Because the distortions of the LF are predominant in the redshift intervals of strong evolution, and because the only redshift interval in which there are enough LF points to make any conclusions about the distortions induced by binning is $0.04<z<0.7$, we choose this redshift interval for our illustration. In Figure~\ref{fig:MOJAVErebin} we present the results of constructing the binned LF using the two methods of sampling with $\Delta \log L = 0.5$ (upper panel) and $\Delta \log L = 0.25$ (lower panel). The continuous line shows the differential model LF derived using a maximum likelihood method \citep[for a description of the procedures and model parameters, see][]{Cara08}. The dashed line shows the binned model LF computed using the method of \citet{PC00} in the $\log L-z$ plane. In order to facilitate the comparison of the distortions of the shape of the LFs we have compensated for the constant bias as described in Section~\ref{sec:MCsim}, that is, we shifted the binned LFs by an amount equal to the bias of the lowest luminosity complete bin computed as $\left\langle \phi \right\rangle_{\mathrm{bin}}-\left.\phi \right|_{\mathrm{bin}\;\mathrm{center}}$. The observed LF binned in the $\log L-z$ plane and represented by filled circles agrees well with the binned model LF constructed using the method of \citet{PC00}, that is, the model provides a good fit to the data. However, both these luminosity functions flatten below $L_{\mathrm{break,1}}$, producing an apparent break. On the other hand, the observed binned LF constructed in the $\log S-z$ plane shows no sign of flattening for $L<L_{\mathrm{break,1}}$. It also confirms our conclusion that the apparent break in the binned MOJAVE LF is due to strong evolution across undersampled bins. The plot in the lower panel was created using a smaller luminosity bin size ($\Delta \log L = 0.25$). One can see that the LF binned in the $\log S-z$ is more tolerant to smaller bin sizes near the flux limit. This is because for luminosity functions with steep (negative) luminosity slopes, the sources will tend to accumulate near the survey limit.

\section{Summary and conclusions}\label{sec:conclusions}

The binned LFs constructed in the traditional $\log L-z$ plane show significant distortions in bins that are undersampled due to luminosity limits of the survey. In the presence of strong positive evolution across these bins, the derived luminosity function will flatten for luminosities below the break luminosities given by equations (\ref{eqn:PosBreak1}), (\ref{eqn:PosBreak2}) and (\ref{eqn:PosBreak3}), producing artificial breaks in the binned LFs. In the case of negative evolution the effect is opposite: the luminosity functions steepen. For single power law LFs such breaks can create the illusion of luminosity evolution. The space density at which these spurious breaks occur varies with the redshift bin.

In order to avoid the possibility of producing spurious breaks in binned LFs, the binning in the $\log L-z$ plane should not extend below luminosities given by $L_{\mathrm{break,1}}$ from equation~(\ref{eqn:PosBreak1}) and binning below $L_{\mathrm{break,3}}$ should be avoided at all costs. However, this may mean that some sources are excluded from the analysis. Alternatively, we propose that the binned LFs be constructed in the $\log S-z$ plane using complete bins only. This method has the following advantages:

\begin{enumerate}
\item Sampling in the $\log S-z$ plane can be done using complete bins without the loss of information (i.e., exclusion of sources with $L<L_{\mathrm{break,1}}$) that would happen if only complete bins were used in the $\log L-z$ plane.

\item Using complete bins eliminates the appearance of spurious breaks or steepening of the LFs.

\item In the same redshift interval the LFs binned in the $\log S-z$ plane can sample the parent LF down to lower luminosities than in the case of $\log L-z$ binning as long as $D^2_{\mathrm{L}}(z)(1+z)^{-(1+\alpharad)}$ is an increasing function of redshift (see equation~\ref{eqn:LcLcDif}).

\item The previous advantage comes at the expense of blurring the real features in the binned LF.


\item Binning in the $\log S-z$ plane should be used in conjunction with binning in the $\log L-z$ plane (for $L>L_{\mathrm{break,1}}$) to test for real breaks in the binned LFs and to identify the features that are the result of binning biases, since the biases associated with binning in the two planes are expected to be different. This is especially important because the binning in $\log S-z$ may blur some weak real features in the binned LF. 

\end{enumerate}

\acknowledgments

We thank the anonymous referee for critical comments that have improved this paper.


\begin{thebibliography}{}

\bibitem[Arshakian et al.(2006)]{Arsh06} Arshakian, T.~G., Ros,
    E.,~\& Zensus, J.~A. 2006, \aap, 458, 397

\bibitem[Avni~\& Bahcall(1980)]{Avni80} Avni, Y.,~\& Bahcall,
    J.~N. 1993, \apj, 235, 694

\bibitem[Cara~\& Lister(2008)]{Cara08} Cara, M.,~\&
    Lister, M.~L. 2008, \apj, 674, 111

\bibitem[Dunlop~\& Peacock(1990)]{Dunlop90} Dunlop, J.~S.,~\&
    Peacock, J.~A. 1990, \mnras, 247, 19

\bibitem[Efstathiou et~al.(1988)]{Efstathiou88} Efstathiou, G.,
    Ellis, R.~S.,~\& Peterson, B.~A. 1988, \mnras, 232, 431

\bibitem[Felten (1976)]{Felten76} Felten, J.~E. 1976,
    \apj, 207, 700

\bibitem[Kochanek (1996)]{Kochanek96} Kochanek, C.~S. 1996,
    \apj, 473, 595

\bibitem[La~Franca~\& Cristiani(1997)]{LaFranca97} La~Franca, F.,
     ~\& Cristiani, S. 1997, \aj, 113, 1517

\bibitem[Lister~\& Homan(2005)]{LH05} Lister, M.~L.,~\& Homan,
    D.~C. 2005, \aj, 130, 1389

\bibitem[Marshall et~al.(1983)]{Marshall83} Marshall, H.~L., Avni,
    Y., Tananbaum, H.,~\& Zamorani, G. 1983, \aj, 269, 35

\bibitem[Miyaji et~al.(2001)]{Miyaji01} Miyaji, T., Hasinger,
    G., Schmidt, M. 2001, \aap, 369, 2001

\bibitem[Padovani et~al.(2007)]{Padovani07} Padovani, P.,
    Giommi, P., Landt, H.~\& Perlman, E. S. 2007, \apj,
    662, 182

\bibitem[Page~\& Carrera(2000)]{PC00} Page, M.~J.,~\& Carrera,
    F.~J. 2000, \mnras, 311, 433

\bibitem[Schmidt(1968)]{Schmidt68} Schmidt, M. 1968,
    \apj, 151, 393

\bibitem[Urry~\& Padovani(1995)]{Urry95} Urry, C.~M.,~\&
    Padovani, P. 1995, \pasp, 107, 803
    
\bibitem[Wall et~al.(2005)]{Wall05} Wall, J.~V., Jackson, C.~A.,
    Shaver, P.~A., Hook, I.~M., \& Kellermann, K.~I. 2005, \aap,
    434, 133

\bibitem[Willott et~al.(1998)]{Willott98} Willott, C.~J.,
    Rawlings, S., Blundell, K.~M.,~\& Lacy M. 1998, \mnras,
    300, 625

\bibitem[Wisotzki (1998)]{Wisotzki98} Wisotzki, L. 1998, AN,
    319, 257

\end{thebibliography}
\end{document}